\documentclass [prl,twocolumn,superscriptaddress,showpacs] {revtex4}
\usepackage{graphicx,amsmath,amssymb,epsf,longtable}

\begin{document}

\title{
Collinear versus non-collinear magnetic order in Pd atomic clusters
}
\author{F. Aguilera-Granja}
\affiliation{Departamento de F\'{\i}sica Te\'orica, At\'omica
y \'Optica. Universidad de Valladolid, E-47011 Valladolid, Spain}
\author{J. Ferrer}
\affiliation{Departamento de F\'{\i}sica.  Universidad de Oviedo, Spain}
\author{A. Vega$^1$}

\date{\today}
\begin{abstract}
We present a thorough theoretical assessment of the stability of non-collinear 
spin arrangements in small palladium clusters. We generally find that 
ferromagnetic order is always preferred, but that antiferromagnetic and
non-collinear configurations of different sorts exist and compete for the 
first excited isomers. We also show that the relative stability of all these states is rather 
insensitive to the choice of atomic configuration for the pseudopotential 
used and to the approximation taken for the exchange and correlation potential. 
This result stands in stark contrast with the situation found for the bulk phases 
of Palladium.
\end{abstract}


\maketitle

The magnetic properties of free-standing atomic clusters of 3d TM 
elements have been intensively scrutinized during the last two decades. 
Two different but related phenomena have specifically been discussed 
and essentially unravelled. The first is the modification of local 
magnetic moments as compared with the values found in bulk materials. 
The second is the competition between the possible ferromagnetic, 
antiferromagnetic and non-collinear arrangements of the local spins, 
as well as its interplay with the geometry of the nanostructure. 
In the case of ferromagnetic elements like Fe,
Co and Ni, the increase of the average cluster magnetic moment can be
easily explained in terms of the reduced atomic coordination in the
low-dimensional regime, with oscillations associated to structural
(symmetry) changes. \cite{fagni}
The case of antiferromagnets like Cr and Mn is much more complex. Atoms 
of these elements may display large magnetic moments, since they have
a large number of d-holes susceptible to be polarized. On the other hand,
clusters of these atoms may display tiny average magnetizations due to the
tendency of their atomic moments to align in antiparallel directions.
The structure plays also a fundamental role in the magnetic behavior of these
clusters, since it may originate magnetic frustration. 
A conventional example of magnetic frustration in a classical spin system appears 
when atoms positions form triangular motifs. The studies of these classical
systems show that magnetic frustration frequently leads to non-collinear 
configurations of the local spin moments.
The latest theoretical studies reported in the literature show that non-collinear 
arrangements of quantum spins also appear as the ground or as some of the first 
isomers of clusters of $3d$ atoms, including not only Cr and Mn, but also Fe, 
Co and Ni.\cite{oda, kohl, hobbs, fujima, longo}

All materials made of $4d$ TM elements are paramagnets, in contrast to those
of the $3d$ row. A natural question thus arises of whether small clusters of 
$4d$ elements may show low-lying magnetic states of collinear or even 
non-colinear nature. Bulk palladium, being a paramagnet in the brink of becoming
a ferromagnet, presents one of the most intriguing and controversial 
magnetic behaviors in nature.\cite{Chen} It is therefore not surprising that 
the very few experimental and theoretical studies published so far try to 
clarify whether Pd clusters of given sizes are magnetic or not, and what is the 
order of magnitude of their average magnetic moment. From the experimental side, 
most of the reports agree that only very small clusters have a net magnetic 
moment \cite{Douglass, Cox, Taniyama, Sampedro}, with the exception of 
Shinohara and coworkers, \cite{Shinohara} who found noticeable magnetic moments 
at the surface of Pd particles as big as 79 \AA. From the theoretical side, there 
is also consensus that very small Pd clusters are indeed 
magnetic.\cite{Lee,Vitos,Reddy,Moseler,Kumar,FAG_Pd}
Futschek et al.\cite{Hafner} have studied recently small Pd clusters 
using Density Functional Theory (DFT) in the collinear framework, 
within a fixed-moment mode.  They have found that multiple spin isomers 
exist for each cluster size with very small energy differences.  Interestingly,
some of these competing isomers present ferromagnetic order, while others display 
antiferromagnetic alignments, with possible frustration. Although Pd has tendency 
to ferromagentic order, this fact strongly points out to
the possible existence of non-collinear magnetic structures, as a mechanism
to release the frustration and competition between the different magnetic solutions.

We report in this article a thorough {\it Ab initio} study of the magnetic 
behavior of small palladium clusters Pd$_N$, with $N$ ranging from 3 to 7.  
We have 
performed a simultaneous optimization of the geometric and magnetic degrees of 
freedom fully allowing for non-collinear spin arrangements. This consists, 
to the best of our knowledge, the first study of non-collinear magnetism 
in $4d$ atomic clusters. 
Moreover, a debate currently exists on the accuracy of the Local
Density Approximation (LDA) \cite{CA} versus the Generalized Gradient Approximation
(GGA) \cite{PBE} for the determination of the magnetic behavior of low-dimensional 
Pd systems \cite{Reddy,Moseler,Kumar,Delin,Alexandre}. The present letter
also assesses the reliability of both approximations for the case of
free-standing Pd atomic clusters.

\begin{figure} \centerline {\includegraphics[angle=-00.0,
    width=0.9\linewidth] {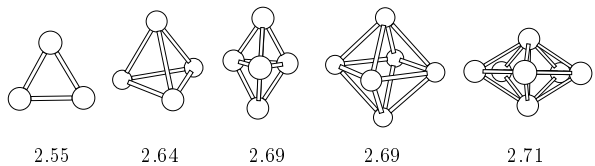}} \vspace{-10pt}
\caption{Illustration of the ground state structures of the different 
clusters here studied and average interatomic distances (in \AA) 
within GGA1.}  \label{Fig1} \end{figure}

We have performed our calculations using the code SIESTA.\cite{SIESTA}
SIESTA is a DFT method that employs linear combination of pseudoatomic
orbitals as basis set. The electronic core is replaced by a nonlocal
norm-conserving Troullier-Martins\cite{TM_1991} pseudopotential that may
include nonlinear core correction terms.  The code allows to
perform, together with the electronic calculation, structural
optimization using a variety of algorithms. It also allows to 
simulate non-collinear spin arrangements both in the LDA and in the 
GGA approximations. \cite{Suarez} 

In the present calculation, we have also used a variety of pseudopotentials
to test their effect on free-standing clusters and their corresponding 
transferability. We have generated three different pseudopotentials 
using LDA. The first (LDA1) was built with the electronic configurations 
$5s^1$, $5p^0$ and $4d^9$, and core-corrections matching radius $r_c=2.00$ a.u.; the
second (LDA2) was identical to LDA1, but with $r_c=1.2$ a.u.; the third
had a closed-shell atomic configuration ($5s^0$, $5p^0$ and $4d^{10}$) and 
$r_c=1.2$ a.u. We have also generated two GGA pseudopotentials with electronic
configuration $5s^1$, $5p^0$ and $4d^9$, and $r_c=2.0$ or 1.2 a.u.
(GGA1 and GGA2, respectively).  In all five cases, the cutoff radii of the s, p 
and d orbitals were taken at 2.30, 2.46 and 1.67 a.u., respectively.
We have described valence states by a double-$\zeta$ polarized basis set (e.g.:
two different radial functions for s and d orbitals and a single one for p orbitals). 
We have taken an energy cutoff of 150 Ry to define the real space grid for numerical 
integrations, but we checked that higher cutoffs did not alter the results.
We have carried out the structural optimization using a conjugate gradient algorithm, 
where we have set the tolerance for the forces at 0.003 eV/\AA, with eventual 
double-checks using 0.001 eV/\AA.

We have found that the five pseudopotentials provide similar results when applied to
an isolated palladium atom, being the eigenvalues of the ground state and
different excited states slightly better reproduced with LDA1 and GGA1 (both had
$r_c=2.00$ a.u.). However, we have observed that they give rise to different magnetic 
behaviors when applied to the bulk fcc material. All LDA approximations give a lattice
constant equal to 3.90 \AA, while all GGA predict it to be equal to 4.01 \AA. 
LDA1 gives a ferromagnetic ground state with $M \approx 0.54 \mu_B$, while LDA2 and LDA3
predict the ground state to be paramagnetic. Finally, both GGA pseudopotentials lead to a 
ferromagnetic ground state with $M \approx 0.48 \mu_B$. These results highlight the importance
of testing all the different pseudopotentials for atomic clusters considered here.  

\begin{table} 
\caption{Bindig energy of the ferromagnetic clusters in meV/atom.}  
 \vspace{-7pt}
\begin{center}
\small{\renewcommand{\arraystretch}{0.10}\renewcommand{\tabcolsep}{0.35pc}
\begin{tabular}{lcccccc}
\hline
 N   & \, LDA1 \,  & \, LDA3 \, & \, GGA1  \, &  \,Ref.[\onlinecite{Kumar}] \, & \,
Ref.[\onlinecite{Hafner}] \,  \\
\hline
3  &   1.755  & 1.326  &  1.289 &  1.203  & 1.250  \\
4  &   2.293  & 1.942  & 1.769  &  1.628  & 1.675  \\
5  &   2.502  &  2.168 & 1.933  &  1.766  & 1.805  \\
6  &   2.721  &  2.401   &  2.110&  1.919  & 1.949  \\
7  &   2.791  &  2.452   &  2.155 &  1.953  & 1.985  \\
\hline
\end{tabular}}
\end{center}
\end{table}

Notice that we have not kept fixed the magnetic moment in our simulations
of the Pd$_N$ clusters, but rather have allowed it to vary freely during 
the non-collinear iterative selfconsistency process, in contrast to previous
authors. Moreover, while we can not rule out that we may have missed
low lying solutions, we have endeavored to minimize this risk
by feeding a large variety of non-collinear seeds for each cluster.
This effort has allowed us to find a rich and complex family of
metastable solutions, that was absent in previous works. We finally
note that we have repeated all calculations with the pseudopotentials
LDA1, LDA3 and GGA1. 

We have found that all clusters, except Pd$_6$, share the same collinear magnetic ground
state, with a total spin of 2 $\mu_B$, in agreement with previous authors \cite{Kumar,Hafner}. 
We should stress that all the tested pseudopotentials provide the same
ground state, in stark contrast to the situation that arose for the bulk
material. Moreover, we have found very similar inter-atomic distances for
all Pd$_N$ clusters, using whichever pseudopotential. These distances also
agree with those obtained by Kumar and Futschek within a range of 1 per cent.
The geometry of the ground state and the average interatomic distance of the
Pd$_N$ clusters is displayed in Fig. 1, where we show that these range from 
2.55 \AA \, in ${\rm Pd}_3$ to 2.71 \AA \, for ${\rm Pd}_7$. We have written the 
binding energies of the different clusters in Table I. The table shows that
GGA1 gives slightly smaller values than LDA1 and LDA3, as otherwise expected. 
Moreover, the binding energies predicted by GGA1 are very similar to those obtained
by Kumar, who also used the GGA (within an ultrasoft pseudopotentials, plane 
waves code) and by Futschek et al., who used the all-electron VASP
code, but did not state the approximation employed. 

\begin{figure}[b] \centerline {\includegraphics[angle=-90.0,
width=0.99\linewidth] {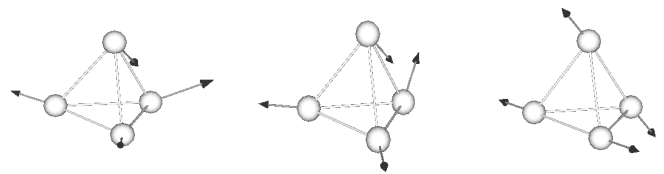}} 
\caption{Illustration of the non-colinear magnetic solutions 
for ${\rm Pd}_4$ NC2 (LDA1), NC2 (LDA3) and NC1 (GGA1). The arrows are proportional 
to the size of the atomic moments.}  \label{Fig2} \end{figure}

\begingroup
\squeezetable
\begin{table*}
\caption{Different solutions obtained for the Pd$_N$ clusters. We provide the absolute
values of the atomic magnetic moments, the total magnetic moment in the cluster $\bar\mu$
(both in units of $\mu_B$) and the excitation energy per atom (in meV). For N=5 and 7, the
first two values of the atomic moments correspond to the axial sites, whereas the last ones 
correspond to the planar sites.}
\medskip
\begin{center}
\small{\renewcommand{\arraystretch}{0.75}\renewcommand{\tabcolsep}{0.001pc}
\begin{tabular}{lcccccccccc}
\hline
\hline
    &LDA1 &  &  & LDA3 &  & & GGA1  & &   \\
\hline
    & Local moments &  $ \bar \mu$ \,\, & $ \Delta E$  & Local moments &  \, $ \bar \mu$ \, & \, $ \Delta E$ \, &
Local moments & \, $ \bar \mu$ \, \, & $ \Delta E$  & \\
\hline
N=3 & & & & & & & & & \\ 
\hline
Ferro. & (0.67$\times$3)& {\bf 2}   & 0 
       & (0.67$\times$3)  & {\bf 2} & 0 
       & (0.67$\times$3)  & {\bf 2} & 0 \\ 
AF     &  &  &   
       &  &  & 
       &  (0, 0.30, -0.30) & {\bf 0} & 28  \\
Radial &  &   &  
       & &  & 
       & (0.18$\times$3) &  {\bf 0} & 28   \\
Para.  &  (0$\times$3) &  {\bf 0}  & 68  
       &  (0$\times$3) & {\bf 0} &  34  
       &  (0$\times$3) & {\bf 0} &  75  \\
\hline
N=4 & & & & & & & & & \\ 
\hline
Ferro. & (0.50$\times$4)& {\bf 2} & 0 &(0.50$\times$4)  & {\bf 2} &0 & 
(0.50$\times$4)& 
{\bf 2} & 0 \\ 
NC1  &  & &  & & &  & (0.29,0.29,0.29,0.29) &{\bf 0} & 12 \\
NC2  & (0.35,0.24,0.24,0.35) & {\bf 0.25} & 9 &(0.25,0.28,0.28,0.25) & {\bf 0.03}  & 10 &&&   \\
AF1  & (0.32, 0.32, -0.32, -0.32) & {\bf 0} & 26 & 
(0,23, 0.23, -0.23, -0.23) & {\bf 0}  & 30 
& (0.29, 0.29, -0.29, -0.29) & {\bf 0} & 25 \\    
AF2 & (0.41, 0, -0.41, 0) &  {\bf 0} & 40 
&  (0,31, 0, -0.31, 0) &  {\bf 0} & 31 
&  (0.38, 0, -0.38, 0) &{\bf 0}  & 36 \\  
Para.    &  (0$\times$4) & {\bf 0} &  86  &  (0$\times$4) & {\bf 0} & 59 &  (0$\times$4) & {\bf 0} & 78 \\
\hline
N=5 & & & & & & & & & \\ 
\hline
Ferro. & (0.43,0.43,0.38$\times$3)& {\bf 2} & 0 
       & (0.40$\times$5)& {\bf 2} & 0 
       &(0.42,0.42,0.39$\times$3)  & {\bf 2} &0 \\ 
AF1 & (0, 0, 0.43, - 0.43, 0)& {\bf 0} & 22 
    &  (0, 0, 0.33, - 0.33, 0)&  {\bf 0}  & 19 
    &  (0, 0, 0.39, - 0.39, 0)&  {\bf 0}  & 18 \\
AF2  &  (0, 0, 0.48, -0.24, -0.24)& {\bf 0} & 27
     &  & &  
     & (0, 0, 0.44, -0.22, -0.22)& {\bf 0} & 19  \\ 
Radial & (0, 0, 0.29$\times$3)& {\bf 0} & 35  
       & & &
       &  (0, 0, 0.27$\times$3)& {\bf 0} & 28 \\ 
Para.      &  (0 $\times$5) & {\bf 0}  & 63  
           &  (0  $\times$5) & {\bf 0}  & 41 
           &  (0 $\times$5) & {\bf 0}  & 55\\ 
\hline
N=6 & & & & & & & & & \\ 
\hline
Ferro. & (0.33$\times$6)& {\bf 2} & 0 
       & (0.33$\times$6)& {\bf 2} & 0 
       & (0.33$\times$6) & {\bf 2} &0 \\  
Para.  &  (0$\times$6) &{\bf 0} & -\,13  
       & (0$\times$6) & {\bf 0} & -\,12   
       &  (0$\times$6) & {\bf 0} & -\,4 \\ 
\hline
N=7 & & & & & & & & & \\ 
\hline
Ferro. & (0.19,0.19,0.32$\times$5)& {\bf 2} & 0 
       & (0.21,0.21,0.31$\times$5)& {\bf 2} & 0
       & (0.20,0.20,0.32$\times$5)& {\bf 2} & 0 \\ 
AF1 & (-0.36,0.36,-0.33,-0.22,0.22,0.32,0)&  {\bf 0} & 9
    &                           &                &     
    & (-0.32,0.32,-0.30,-0.20,0.20,0.30,0)&  {\bf 0} & 8   \\ 
AF2 &  (0,0,-0.36,-0.23,0.23,0.36,0)& {\bf 0} & 14
    & \, (0,0,-0.29,-0.20,0.20,0.29,0)& {\bf 0} & 8  
    &  (0,0,-0.32,-0.21,0.21,0.32,0)& {\bf 0} & 12   \\
Radial  & (0.27,0.27,0.18$\times$5) & {\bf 0} & 22   
        & (0.24,0.24,0.12$\times$5) & {\bf 0} & 14 
        & (0.24,0.24,0.17$\times$5) & {\bf 0} & 20 \\
Para.   &  (0$\times$7) & {\bf 0} & 37  
        & (0$\times$7) & {\bf 0} & 24  
        & (0$\times$7) & {\bf 0} & 33 \\ 
\hline
\hline
\end{tabular}}
\end{center}
\end{table*}
\endgroup

The Pd$_6$ cluster displays a behavior different from the rest, and therefore we discuss it
separately. Futschek and coworkers \cite{Hafner} found 
that Pd$_6$ was also ferromagnetic in contrast to Kumar et al.\cite{Kumar},
who predicted it to be paramagnetic. We have found that both states are
nearly degenerate, with the paramagnetic solution being slightly more
stable. Aditionally, we have been unable to find non-collinear or antiferromagnetic
solutions for this cluster.

In contrast, and independently of the pseudopotential or approximation used, 
the rest of the clusters show a rich variety of antiferromagnetic and non-collinear 
solutions. Most of these solutions, though not all, exist for all LDA1, LDA3 and GGA1.
We have also found that, whenever they exist, the relative order of the different
solutions is maintained, and the size of the atomic moments is very similar. These facts 
strengthen our belief that Pd atomic clusters are much more insensitive to the 
pseudopotential and approximation employed than bulk Pd. It is also reassuring
that most of the collinear solutions have been identified in previous 
calculations\cite{Hafner} (e.g.: AF1 for Pd$_4$ and Pd$_5$ and AF2 for Pd$_7$).

\begin{figure} \centerline{\includegraphics[height=7cm,width=5cm,angle=-90]{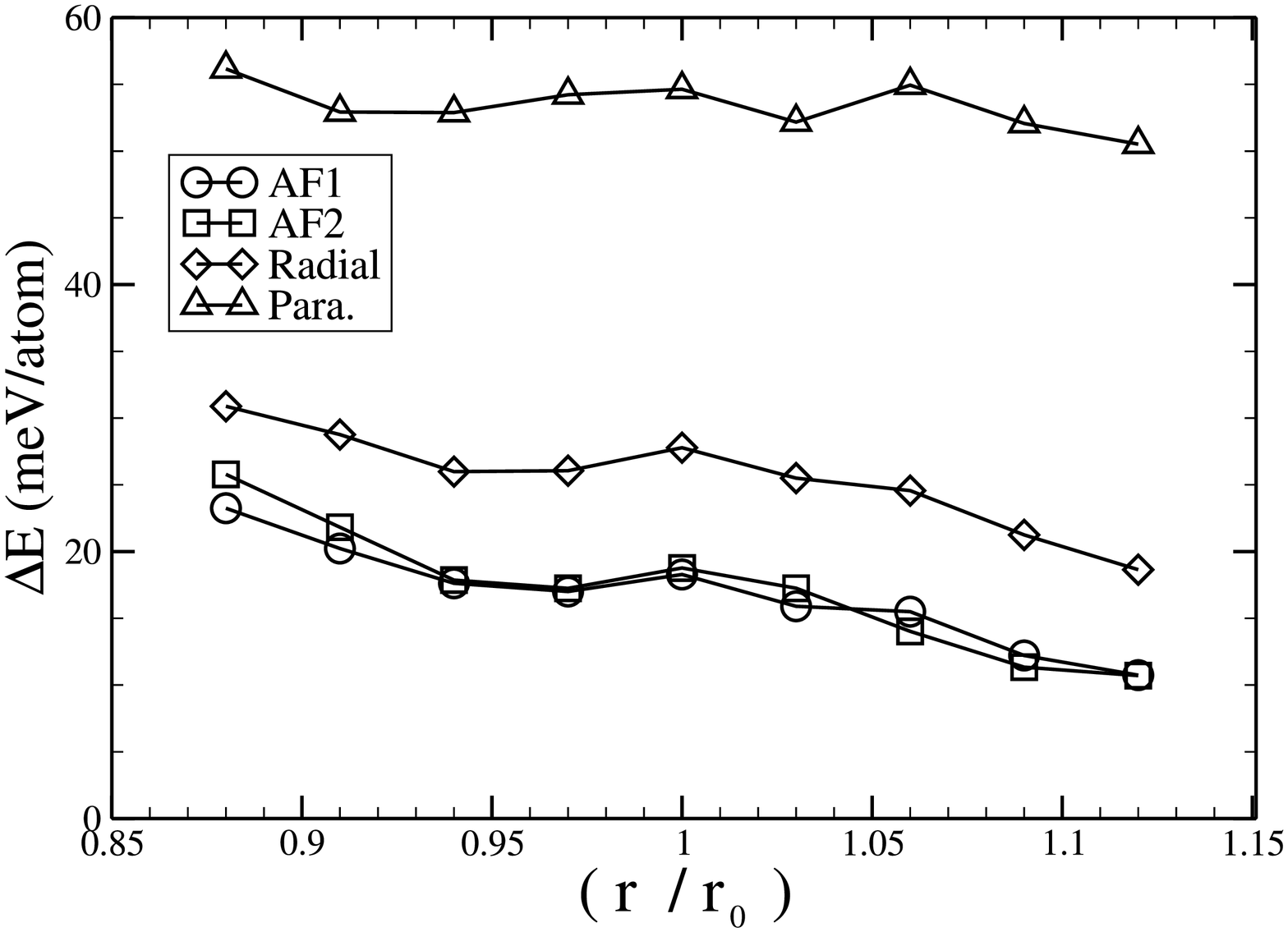}}
\caption{Excitation energy per atom of the magnetic solutions of the Pd$_5$ cluster as a 
function of the average interatomic distance, using GGA1.} \label{Fig3} \end{figure}

The non-collinear solutions found can be classified into those that release antiferromagnetic
frustration and therefore have lower excitation energy than the AF solution
(NC1 and NC2 in Pd$_4$, shown in Table II and Fig. 2), and radial or quasi-radial
solutions, that resemble the hedgehogs found in low dimensional theories of
classical or quantum antiferromagnets\cite{Haldane}. Hedgehogs in these theories
do not release frustration but rather are excitations over the antiferromagnetic ground 
state. We also find that these radial states have a higher energy that the antiferromagnetic
solution, and therefore do not release frustration.

Notice that the antiferromagnetic and non-collinear solutions can be
reached at temperatures of the order of room temperatures (25 meV).
Therefore, any measurement of the magnetization performed at room 
temperature should find a thermal average of all those states, many 
of which have a tiny magnetic moment. It should not be surprising that
such a measurement give a small net moment. 

We finally discuss the relationship between magnetism and equilibrium interatomic
distances. We have found that these are essentially the same regardless of
the magnetic state for the largest clusters (n = 5 - 7), the smallest ones
showing slight variations of less than 0.04 \AA, but only within the
LDA solutions. We have additionally analyzed the relative stability of the 
different solutions as a function of the interatomic distance. To this aim, 
we plot the energy per atom of the low-lying excited states of the Pd$_5$ cluster, 
relative to the ground state energy, as a function of an uniform volume expansion,
obtained using GGA1. The figure shows that no crossover takes place, apart from
the nearly-degenerate AF1 and AF2 solutions, that cross at an expansion of about 
4\%. Moreover, the relative energy differences are essentially preserved and the 
local magnetic moments kept constant, except for the AF2 and radial solutions,
where they slightly change (by about 10\%).

To summarize, we have studied the geometry and magnetic properties of the ground 
state and lowest lying isomers of small palladium clusters Pd$_N$, with $N$ ranging 
from 3 to seven. Our results confirm that the ground state is indeed collinear or 
paramagnetic. We have found a rich variety of non-collinear low-lying isomers, some 
of which efficiently release frustration, while other (hedgehog-like solutions) do not.
All these solutions should contribute to the room temperature magnetic behavior of the 
clusters, probably rendering small measured magnetic moments.
We have finally found that all these states are rather insensitive to the
choice of the pseudopotential and to the approximation
used for the exchange and correlation potential.

This work was supported by the Spanish Ministerio de Educacion y Ciencia 
(grants MAT2005-03415, BFM2003-03156 and SAB2004-0129), INTAS (Project 03-51-4778)
and the Mexican (PROMEC-SEP-CA-230 and CONACyT-SNI). We wish to acknowledge useful 
conversations with J. M. Montejano-Carrizales, L. Fern\'andez-Seivane and F. Yndurain. 

{}

\end {document}